# INFORMATION SECURITY SYNTHESIS IN ONLINE UNIVERSITIES

MARIA SCHUETT, CRISC[1] AND SYED (SHAWON) M. RAHMAN, PHD[2]

[1]Information Assurance and Security, Capella University, Minneapolis, MN, USA
MSchuett@capellauniversity.edu
[2]Assistant Professor, University of Hawaii-Hilo, HI, USA and Adjunct Faculty, Capella University, Minneapolis, MN, USA
SRahman@Hawaii.edu

## ABSTRACT

*Information assurance is at the core of every initiative that an organization executes. For online universities, a common and complex initiative is maintaining user lifecycle and providing seamless access using one identity in a large virtual infrastructure. To achieve information assurance the management of user privileges affected by events in the user's identity lifecycle needs to be the determining factor for access control. While the implementation of identity and access management systems makes this initiative feasible, it is the construction and maintenance of the infrastructure that makes it complex and challenging. The objective of this paper[1] is to describe the complexities, propose a practical approach to building a foundation for consistent user experience and realizing security synthesis in online universities.*

## 1. INTRODUCTION

Sustaining confidentiality, integrity and availability of information assets is a common objective for achieving information assurance. It is an ongoing effort that necessitates thorough understanding of the risks surrounding authentication and authorization of information assets. In a traditional brick and mortar university access control is managed with picture IDs. The amount of times that a student must identify himself is kept to a minimum, for instance, during class registration, or at the start of the semester. Once that initial face-to-face identification is completed at the beginning of the semester, the student may never be asked to show his identification again. Seamless access to resources like the bookstore, laboratory, registrar's office, library, recreation hall, or auditorium, is available to the student. Proving one's identity is kept to a minimum. It's a lifestyle and an environment that enables the student to easily adapt.

While online universities intend to provide the same type of seamless experience, the complexities start with laying the foundation for identity and access. How does an online university provide seamless access and consistent user experience similar to a traditional brick and mortar university? A practical approach is to use role management as the groundwork of access rights management. The identity lifecycle of a student incorporates many roles affecting access privileges to many resources. Apart from managing student lifecycles, is the management of risks that surround the identity and access management infrastructure. How should the IAM infrastructure be secured so that it can function efficiently and effectively? This paper will use

---
[1] This work is partially supported by EPSCoR award EPS-0903833 from the National Science Foundation to the University of Hawaii.

Alpha Educational Institution as a case study to describe typical challenges and practical approaches for online universities.

## 2. OVERVIEW OF THE ORGANIZATION

Alpha is an online educational institution providing undergraduate degrees and higher education to students who aspire to pursue Graduate, PhD, and Certificate programs. Most students enroll at Alpha to pursue higher education. It serves over seventy five thousand students, employs over one thousand eight hundred administrative staff at its headquarters, and over fifteen hundred faculty members around the country. There are two types of users at Alpha, the remote users and the onsite users. Majority of the remote users are the students and faculty. Staff or employees are mostly onsite users.

At Alpha, the management of security is a hybrid of centralized and decentralize administrative services ranging from user account creation, password resets, and authorization to access resources. Resources incorporate UNIX and Windows servers, network components, infrastructure applications, business applications, and databases. Infrastructure applications include user registries (e.g. Active Directory, LDAP), and Identity Management (IDM) and Access Management (AM) systems. Business applications include PeopleSoft HCM, PeopleSoft SA, PeopleSoft Finance, and Blackboard Learn, etc. The hybrid approach of managing security has been the standard practice since Alpha's early years when it employed less than seven hundred employees and served less than ten thousand students.

The IT environment consists of non-production and production environments that support the overall hardware and software lifecycle management infrastructure at Alpha. Development, Quality Assurance, and User Acceptance Test are the non-production regions. Software development lifecycle for new projects and modifications to existing systems or applications go through all three non-production regions (i.e. Development, Quality Assurance, and User Acceptance Test) prior to deployment into Production. Alpha utilizes change management control when deploying changes into Production. Each region consists of its own firewalls, UNIX and Windows server, infrastructure applications, and business applications.

## 3. DEFINITION OF NEED

Alpha uses PeopleSoft HCM/SA system for enterprise resource planning, along with Identity and Access Management systems for managing identities and controlling access. Alpha Educational Institution is a merchant who accepts and processes payment cards; therefore, complying with Payment Card Industry (PCI) Data Security Standard (DSS) is essential. Enhancing the security of the identity management system necessitates understanding the requirements of PCI-DSS and Family Educational Rights and Privacy Act (FERPA). FERPA is a Federal law that protects the privacy of student education records.

The PCI DSS requirements include developing and maintaining secure systems and applications; encrypting transmission of cardholder data across open, public networks; protecting stored cardholder data; and assigning a unique ID to each person to access resources within the organization [1]. The FERPA regulation entitles a student the rights to access to his or her education records, and have control over the disclosure of personally identifiable information (PII) [2]. Both the PCI DSS standard and the FERPA regulation will be used as guidelines to deliver a practical security strategy.

A high level illustration of the identity and management system is shown on Figure 1. Designing and customizing the identity management system requires knowledge of how and when data flows into the organization, where data it is stored, and how it is used. PeopleSoft

HCM/SA is the authoritative source of Alpha Educational Institution. The identity management system pulls new user information from PeopleSoft HCM/SA and creates an account for this user in the identity management system. Consequently, the identity management system automates provisioning and de-provisioning of resources to all users across the organization. It manages the user's lifecycle events and provides password synchronization for efficient access to resources. Events are changes in the user's status that affect his or her access rights.

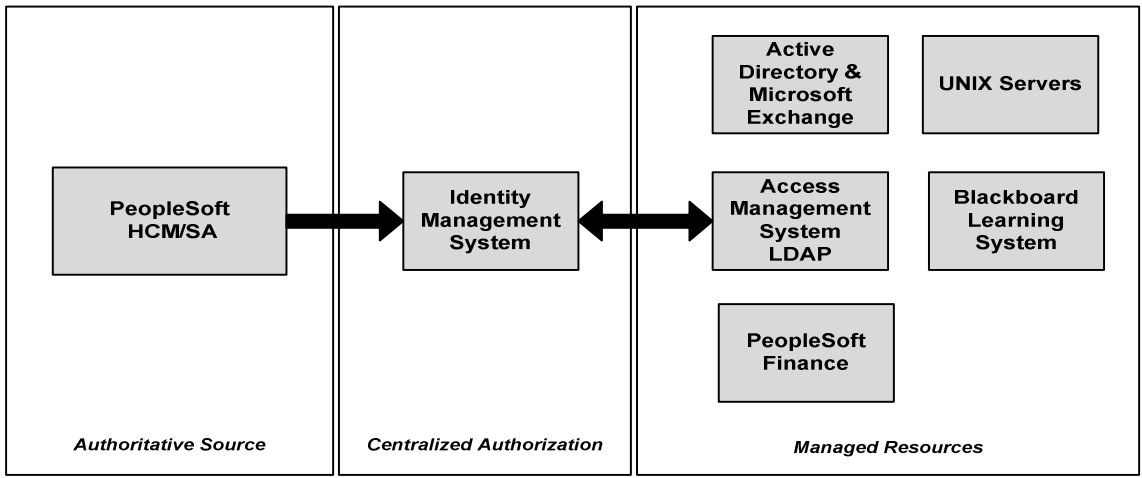

Figure 1. Identity and access management system [3]

Recognizing that employee events differ from student and faculty events adds complexity to managing access rights. For instance, when an employee goes on a medical leave for a number of months, what changes need to be implemented to control and protect the employee's entitlements? Should the employee's account be disabled or left enabled? Or when an employee moves to another department, what should be done to his or her previous access rights to resources? For the most part, that is manageable. Compared to an employee, a student's lifecycle in an online university incorporates more events. A few examples of events are: application, matriculation, enrollment, withdrawn, graduation, and alumni. Whatever the case may be, all user events require the identity and access management systems to accurately maintain a user's access rights. By and large, the identity and access management systems are used to manage access and enforce security policies through a common infrastructure.

## 4. PROPOSED SECURITY APPROACH

Alpha Educational Institution desires to enhance technology, and processes to have better control over the entire organization. In recent years, the number of employees, faculty, and students has more than doubled. While it has managed to maintain hybrid security architecture, when the organization was much smaller, Alpha now requires centralized security architecture as depicted on Figure 1. The following are strategic recommendations:

1. Define consistent and repeatable processes
2. Enhance role-based access to information assets
3. Centralize access control for applications

### 4.1. Define Consistent and Repeatable Processes

The evaluation of practices, processes, and current technology across the organization are steps required to gain an overall understanding of identity management. To further illustrate, processes need to change to support the automation that IDM provides when provisioning or de-provisioning managed resources. The discovery of processes not only includes documented processes but should also include the thought process that one assumes when making access rights decisions. This logic along with approval/rejection processes need to be incorporated into the automated processes. Defining consistent and repeatable processes means that steps have been refined to be efficient and ensure security.

## 4.2. Enhance Role-Based Access to Information Assets

An identity management system's primary function is to automate user management which includes account creation, account updates, and account deletions on various resources. Resources incorporate UNIX and Windows servers, network components, infrastructure applications, business applications, and databases. Currently, Alpha Educational Institution uses role-based access control to authorize users to various resources applications. As described on Figure 1, PeopleSoft HCM/SA is the authoritative source of new identities and all updates to identities. The authoritative source is configured to determine the new user's role to be one of the following: a student, an employee, faculty, or contractor. The identity management system pulls new user information from PeopleSoft HCM/SA and creates an account for this user in the identity management system. Specifically, IDM provisions resources to this user, based on predefined authorization rules related to the user's role. For instance, if this user's role is an employee then IDM will provision Active Directory and Microsoft Exchange, and LDAP. To illustrate, a simple high level provisioning approach would look like Table 1. To enhance role-based access to information assets, the IDM system should be designed to automate user provisioning based on this table.

Table 1. Provisioning table. [4]

| Role/ Resources | Access Management System (LDAP) | Active Directory & Microsoft Exchange | Unix Servers | Blackboard Learn |
|---|---|---|---|---|
| Employee | X | X | X | |
| Student | X | | X | X |
| Faculty | X | | X | X |
| Contractor | X | X | | |

While role management is straightforward for managing employees, it becomes elaborate when it comes to managing a student's lifecycle. The natural progression of an online student starts with the following events.

1. Application event process
2. Matriculation event
3. Enrollment event
4. Graduation event

The complexity of role and entitlement management is between enrollment and graduation. Graduation may not always be the last event, there are events to consider. For instance, if the university withdraws a student for academic failure, then that is an event to manage. If the student decides to withdraw for financial reasons, that is another event to manage. Furthermore, access to resources does not have to end after graduation, or when the student's role changes to alumni. There are many thought processes involved in making these events transparent in order to automate IDM to provision, and de-provision user accounts. The overall goal of information assurance is the management of access rights which are affected by events in the student's identity lifecycle as it is the determining factor for controlling access to resources.

To summarize, the process of automating IDM services using workflows should be deployed to the non-production and production regions to ensure consistency of role-based provisioning and de-provisioning across all environments. The provisioning table provides a pictorial guide of increasing the amount of resources that can be provisioned by IDM. The recommendation is for Alpha to further categorize the employee and student roles so that specific resources can be assigned to these roles, with examples shown in Table 2. Overall, adding more roles and resources helps pave the road to centralizing application access using delegated administration.

Table 2. Additional roles. [5]

| Role/ Resources | Access Management System (LDAP) | Active Directory & Microsoft Exchange | Unix Servers | Student Portal | Blackboard Learn |
|---|---|---|---|---|---|
| Employee | X | X | X | | |
| ✓ Management | X | X | X | X | |
| ✓ Individual Contributor | X | X | X | | |
| Student | X | | X | | X |
| ✓ Active | | | | X | X |
| ✓ Inactive | | | | | |
| ✓ Alumni | | | | X | |
| Faculty | X | | X | | X |
| Contractor | X | X | | | |

## 4.3. Centralized Access Control for Applications

The access management system, on Figure 1 works in conjunction with IDM to manage access to information assets. Access Manager (AM) is a framework that is designed to provide authentication and authorization by centrally controlling application access and improving user experience through single sign-on (SSO). At Alpha, the applications protected by AM are Active Directory, Microsoft Exchange, and Blackboard Learn.

The PeopleSoft applications namely, Human Capital Management (HCM), Student Administration (SA), and Finance, house Personally Identifiable Information (PII). PII includes full name, maiden name, social security number, street address, telephone numbers, and other pieces of information that can positively identify a person [6]. Alpha Educational Institution should enhance the security of the PeopleSoft systems, as mandated by PCI DSS and FERPA. The following steps will help Alpha Educational Institution achieve the requirements of PCI DSS and FERPA.

1. Secure the PeopleSoft applications behind AM to strengthen authentication.
2. Implement PKI for accessing PeopleSoft, and Blackboard Learn applications to correspond to PCI standards and FERPA regulation.
3. Centralize authorization by using AM to enforce fine-grained grained access control to PeopleSoft, and Blackboard Learn applications.

Realizing that the scope of access control is not merely letting the administrator and the end-user into the application is vital to providing fine-grained authorization. Within each application, whether it is PeopleSoft HCM/SA or Blackboard Learn, are many capabilities of access control and authorization. So understanding the applications' access and authorization capabilities along with the enterprise Access Management system can provide a depth of defense security controls for the online university.

### 4.3.1. Public Key Infrastructure

An important step in centralizing authentication for PeopleSoft applications, using AM, is to implement PKI for accessing PeopleSoft applications, and Blackboard Learning System to correspond to PCI standards and FERPA regulation. Public Key Infrastructure (PKI) enables entities to communicate securely in a trusted manner using public key cryptography and digital certificates using the X.509 standard. An entity is a person, an organization, a web server, or an application server. These entities need to communicate in a trusted and secure manner. Public key cryptography addresses security by providing confidentiality, authentication, and non-repudiation with the use of a private key and a public key that are linked mathematically, and mutually exclusive. Both keys can be used to close or open the lock, however, the same key cannot be used to do the same function. So if the private key was used to encrypt the message then the public key is needed to decipher the message. If the public key was used to encrypt the message then the private key is needed to decipher the message. However, using public key cryptography alone without the assurance of trust is a weakness. PKI resolves the weakness with the use of the X.509 certification infrastructure. The X.509 is a standard for PKI [7].

PKI components include a registration authority (RA), and a certificate authority (CA). The function of the RA is to accept the entity's certificate registration request. The entity's certificate request includes unique attributes that identify the organization. Once the RA has verified the authenticity of the entity, the request is forwarded to the CA. Only the CA can generate the digital certificate using the X.509 standard. The CA is a trusted authority which can be internal or external to the organization. When the PKI does not include an RA, then the primary role of a CA is to validate the identity of the entity requesting a digital certificate. In essence, the CA and the RA's function in authenticating the entity is the first step to building trust. The second step to establishing trust is with the use of digital certificates, as it indicates the entity's signature of authenticity with the CA as the authorized facilitator. "One of the most important pieces of a PKI is its digital certificate. A certificate is the mechanism used to associate a public key with a collection of components in a manner that is sufficient to uniquely identify the claimed owner [8]." The X.509 certificate standard provides interoperability and consistency by using formats for and attributes of public key certificates in a hierarchical trust model [9].

The role of a CA is to create digital certificates, digitally sign, and deliver digital certificates, and maintain the digital certificates over its lifetime. A certificate revocation list (CRL) is used to maintain a list of invalid digital certificates that have either expired or compromised. The successful maintenance of digital certificates requires trust between the entity and the CA.

By assuring trust between the participating entities, these entities can communicate across networks, across countries, around the world securely. Trust is the confidence that entities are communicating securely to each other in a controlled environment. PKI is the best choice for securing organizational infrastructure and in SDLC because it provides the foundation to build confidentiality, and integrity within the IT infrastructure. It controls access to information assets by authentication and authorization. PKI ensures the organization's accountability with non-repudiation. Trust is inherent with the use of a PKI because it can disguise and protect data and when incorporated into the SDLC process.

The following diagram show on Figure 2 is how PKI can be used in the authentication process.

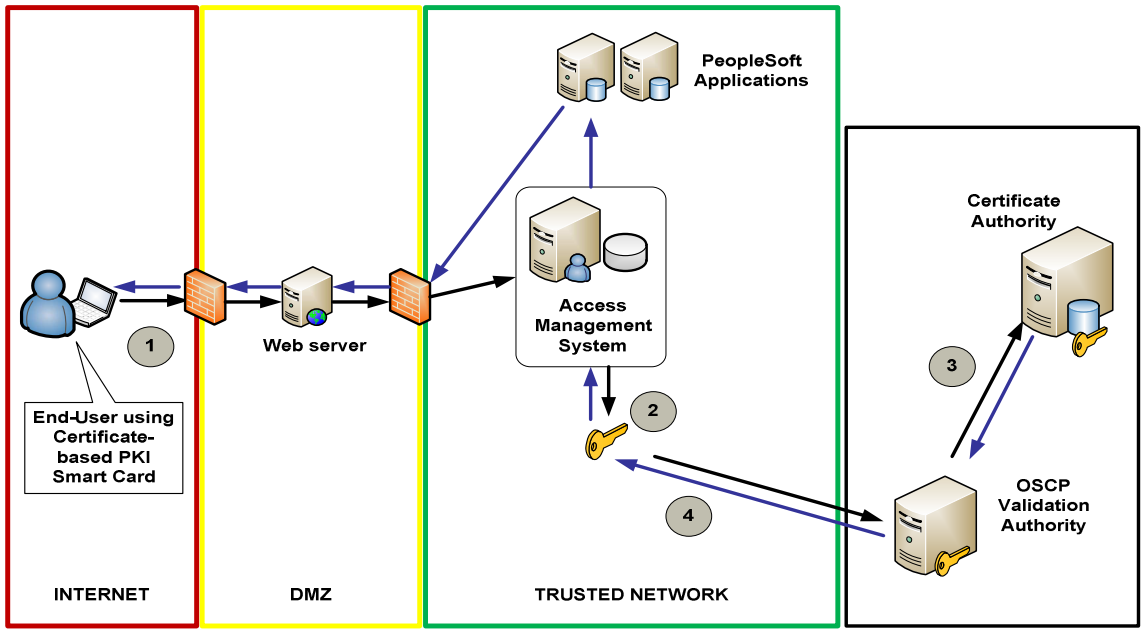

Figure 2.  Authentication in a PKI environment [10]

At Alpha, the types of users that access the PeopleSoft applications are either remote or onsite. For coherent access control, the recommended approach is to use PKI which will provide the framework for multi-factor authentication.   This approach uses extensible authentication protocol (EAP) [11] to deploy a common type of authentication method used in wireless networks and point to point connections.  It has the capability of supporting multiple types of authentication methods such as id and password, tokens, digital certificates, smartcards and biometrics.  To ensure the integrity of the authentication method, AM will be configured to use an Online Certificate Status Protocol [12], instead of a Certificate Revocation List (CRL). OSCP alleviates the challenge or commitment needed to frequently update the CRL list.  Figure 2 depicts a remote user using multi-factor authentication to access PeopleSoft applications.  An on-site user would follow the same scheme when accessing PeopleSoft applications in the Production region. Figure 2 is the recommended approach for the production environment.

1. The end-user, who is offsite, presents own public-key to access PeopleSoft applications
2. AM determines the revocation status of the certificate by issuing a real-time status request to the OSCP Validation Authority or Certificate Authority
3. OSCP Validation Authority or Certificate Authority validates the end-user's public-key with the corresponding private key
4. AM confirms the status by accepting the response from the OCSP responder.  The end-user is successfully authenticated and can access PeopleSoft Applications

To better secure non-production environments the recommended approach is to use secure HTTP and deploy certificate based authentication in addition to forms authentication.  The same scenario would apply as depicted on Figure 2 with the exception of using LDAP instead of OSCP.

1. The end-user presents own public-key to access PeopleSoft applications
2. AM determines the validity of the PKI certificate by authenticating the end-user's public-key against the user's LDAP account

3. AM uses the X.509 attributes from the certificate (i.e. UID, emailAddress, etc.) to search and retrieve the stored end-user's certificate from LDAP
4. If the end-user's certificate matches the retrieved certificate, successful authentication is achieved and the end-user can access PeopleSoft Applications

These two approaches for remote authentication and on-site authentication should be deployed for other core applications (i.e. Blackboard Learning System) protected by the access management system. This configuration helps protect PII and FERPA information by keeping the data confidential and only accessible to authorized users on a need-to-know basis.

From a system administrator's perspective, using AM to centrally enforce fine-grained access control to PeopleSoft applications, and Blackboard Learning System, will provide the organization transparency of the strengths and challenges of application administration. Using delegated administration provides a basis of improving access control and enables the organization to generate audits and reports for compliance to security policies, PCI DSS, and FERPA. Enforcing fine-grained access control entails a security team owning the administration of AM, and delegating application control to the application owners or application custodians. Table 3 shows an example of delegated administration for managing fine-grained application access.

To recapitulate, securing the PeopleSoft applications behind the access management system is a way to strengthen authentication and ensure against malicious ciphers. With PKI as the underpinning of the identity and access management systems, it is becomes possible to securely manage the technology, the people, and the processes.

Table 3. Using delegated administration for fine-grained access. [13]

| Action/Role | AM Domain Admin | Senior Application Admin | Application 1 Admin | Application 2 Admin |
|---|---|---|---|---|
| Manage application groups | X | | | |
| Add member to group | | | X | X |
| Modify user's access | | | X | X |
| Delete member from group | | | X | X |
| Create/View groups | X | X | | |
| Create/View sub groups | X | X | | |
| Assign application admin | X | | | |
| View members | | X | X | X |

## 5. SECURING IDENTITY AND ACCESS MANAGEMENT

Securing the IAM infrastructure requires the evaluation of all related components. It incorporates the network, the operating system, the user registry or databases, and the IAM configuration, including data encryption. The following sections will describe specific areas to enhance for better security,

1. Identity and access management configuration
2. Database security

## 5.1. Identity and access management configuration

This objective is to ensure that the identity and access management infrastructure components systems are configured to communicate in a secure manner to preserve the integrity, confidentiality, and availability of information assets. At Alpha, the components that need to be protected within the identity and access management realm are application servers, database servers, web servers, load balancers, and firewalls. The communication channel or transport layer between each component should be secured using Secure Socket Layer (SSL) protocol. To achieve non-repudiation, the creation of SSL certificates should be done using Public-Key cryptography.

In Figure 3, the access management system communicates with the LDAP server to validate a user's credentials. Since this channel contains users' credentials it should be protected from vulnerabilities using SSL certificates. Similarly, Blackboard Learn and PeopleSoft HCM/SA applications use databases to hold PII information so securing the transport layer with SSL is required to ensure:

1.  That the only entity allowed to write to and read from the database is the application itself
2.  The prevention of other entities or programs from querying the database with the use of mutually authenticated certificates between the web servers and the applications (i.e. PeopleSoft HCM/SA & Blackboard learn)

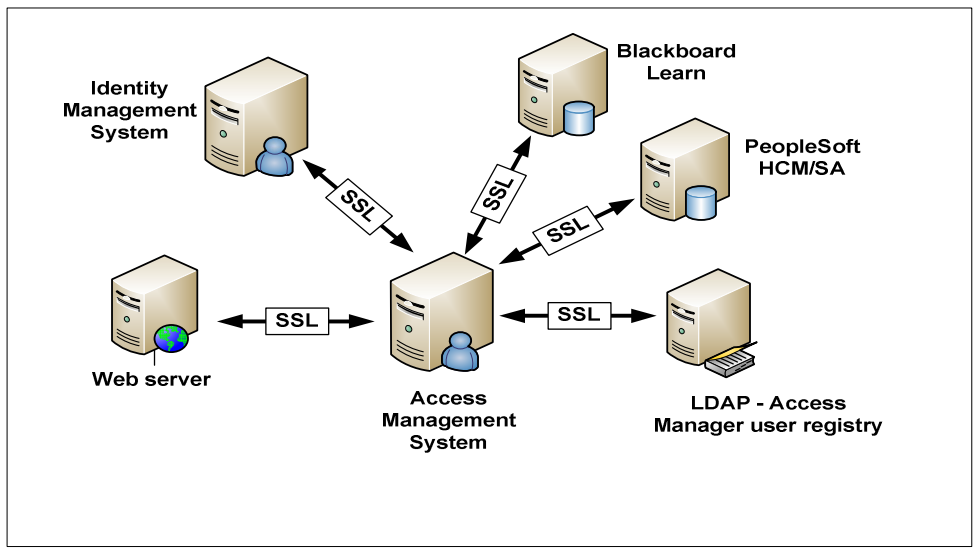

Figure 3. Secure communication using SSL certificates [14]

## 5.2. Database Security Considerations

The evaluation of database security controls explores the configuration between the operating system, application, and database, as depicted on Figure 4. Making sure that only appropriate components are integrated together prevents unauthorized changes or access to the data.

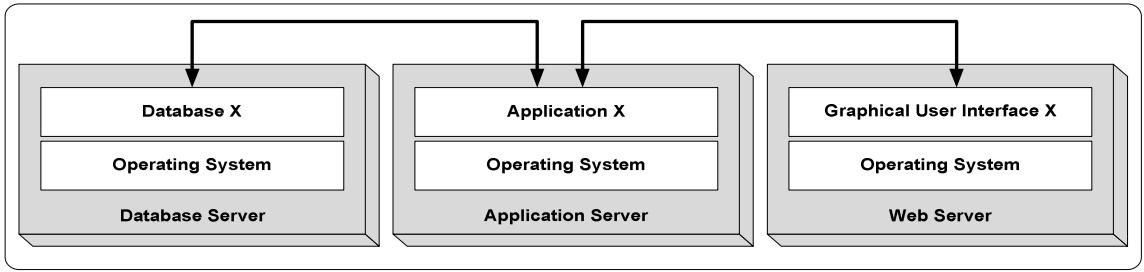

Figure 4. 3-Tier application infrastructure [15]

At Alpha, the management of the application infrastructure involved many roles. Table 4 describes how data is created, where data is kept, and how data evolves within the organization

Table 4. Categorization of data. [16]

| Type of Data | Description |
|---|---|
| Electronic University Student Data Repository | Data derived from electronic systems regarding current students, prospective students. Blackboard Learning System is in this category. |
| Integrated Student Information System | These systems store student information and data - PeopleSoft SA, PeopleSoft Customer Management System (CRM), Electronic Document Management System, and Identity Management System. |
| Shared Student Data Repository | Any data repository that is used by multiple organizations internal to Alpha Educational Institution to achieve various objectives. This includes the Access Management System that uses LDAP as a user registry. |

The types of job functions are directly related to the maintenance of the data describe in Table 4 and can be categorized in the following:

1. The administrators (i.e. application, database, system, network)
2. The end-users (i.e. developers, customers)

The administrators are responsible for operating system, database, and application installations. The end-users are the developers who add functionality to the application, and the customers are the students, faculty, and employees of Alpha. From a security perspective, Table 5 describes the roles and responsibilities of maintaining data within the organization.

Table 5. Roles regarding the use, and maintenance of data. [17]

| Roles | Responsibilities |
|---|---|
| Data Providers | Students, faculty, employees, and contractors are data providers, as end-users. |
| Data Owner | Responsible for the security and use of the data. This role can be a senior executive, or organization leader, held accountable for the protection of information assets. |

| Data Steward | The steward represents the data owners by making decisions regarding security policies, and architecture. This role manages the integrity, privacy, and accuracy of the data. |
|---|---|
| Data User | This user is the consumer of the data or the end-user. Any authorized user that utilizes university data to perform their jobs. This role has the responsibility of working with the data custodians, and data stewards, manage, and secure the data. |
| Data Custodian | This role manages the security of all data for the data owner. Securing data means preventing un-authorized access and un-authorized use, maintaining the integrity and availability. The administrators (i.e. application, database, system, and network) collaborate with the data owner and data users to implement preventive, detective, and corrective security controls. |

In this case study, the critical applications that utilize databases are IDM, PeopleSoft HCM, PeopleSoft SA, and Blackboard Learning System. All these systems follow a basic application and database layout, as shown on Figure 4. The evaluation of the security architecture surrounding database security and secure application development requires the following controls and to be analyzed.

1. Database security considerations
2. Software development lifecycle (SDLC) and its effect on revision control of secure systems

### 5.2.1. Database Security Considerations

One security consideration is the use of database links. "Database links are one-way communication path from an Oracle database to another [18]." Public links are more risky than private links. Public links allow all users and programs access to a remote database, once authenticated on the local database. Controlling access using public links is a challenge because the link contains the user's credentials. Therefore, anyone who can see the link will have access to the remote database, and worse the audit logs on the remote database will not show the user's actions. Private links only allow the owner access to the remote database. Overall, avoid using public and private links because the credentials are stored in clear text in a table.

It is important to understand the internal and external business drivers, at Alpha, so that security policies and standards can be created and enforced consistently in both non-production and production environments. This includes establishing trust relationships Trust relationships between databases and operating systems enable data custodians administer the database once authenticated at the operating system lever. Some organizations use trust relationships between databases and operating systems, by way of single sign-on [19]. This is defined as pass-through authentication for an external user. Establishing trust between the database and the local host or operating system is a better practice than trusting a remote host to authenticate the user. To mitigate this risk, avoid trusting a remote host to perform authentication for an external user. Trusting internal users mean assigning the account to a group that can administer the database. This group is maintained by the operating system. This practice lacks audit trails because all members of the group can access the same internal account [20]. Mitigating this risk means ensuring that only authorized users are in the group, and monitoring usage.

The following diagram, Figure 7, shows the graphical user interface dedicated to user authentication. Bypassing this control is not recommended when developing database applications. Trusting other entities to perform authentication for the database is also a common practice that needs better control to prevent identity spoofing. The next security consideration is the use of dynamic SQL. Dynamic SQL is a programming methodology for generating and executing SQL statements at run time. It is used as a general-purpose program for flexible hoc queries. It is dynamic in a way that it can extract any type of information from the database with user inputs directly in SQL query. It can either execute immediately or gather the information and then execute. The use of dynamic SQL should be avoided because, if compromised, it can expose sensitive data to un-authorized users. It may expose confidential data, and compromise the integrity, and availability of the database. Dynamic SQL enables SQL injection attacks. SQL injection attacks can come into the network through HTTP through an open port on the firewall, usually port 80. Recommended mitigation is to use static SQL, which has the following advantages

1. Have a way to control the user's input by "bind variables" or "parameterized SQL." Instead of directly inserting user-supplied input into the SQL query, inputs are first assigned to parameters (variables) and then the parameters are user in the SQL query [21].
2. Successful compilation verifies that static SQL statements reference valid database objects and that the necessary privileges are in place to access those objects.
3. Successful compilation creates schema object dependencies.

### 5.2.2. SDLC and its effect on revision control of secure systems

Software development lifecycle (SDLC) is a set of guidelines that help an organization analyze, design, develop, deploy, and maintain commercially purchased and in-housed developed software.

When developing database applications or installing third party vendor software, it is important to incorporate the SDLC process to reach security objectives. SDLC involves the following phases with security as the core of each phase.

1. Initiation – define security requirements, risk assessment
2. Acquisition/Development – define initial baseline of security controls to mitigate identified risks, security control design, security planning
3. Implementation/Assessment – security control integration, security certification and accreditation
4. Operations/Maintenance – change control, incident handling, auditing, intrusion detection, and monitoring
5. Sunset or Disposition – transition planning, component disposal, information archiving

Security in SDLC, as shown on Figure 5, can be used as a guide to achieve database security and promote secure application creation [22]. One important function of the SDLC process is to promote the use of change control management. It helps formalize a process of deploying new applications or modifications from the development to the production environments.

The SDLC model incorporates the use of database encryption, and key management to ensure confidentiality and integrity. Preventive controls ensure that database replication and backups are effective. Detective controls monitor intrusions and un-authorized access through logging. Overall, the use of identity and access management systems, in combination with SDLC will help Alpha achieve its security objectives. The next section will clarify software quality, reliability, and security to provide more insights about application development practices.

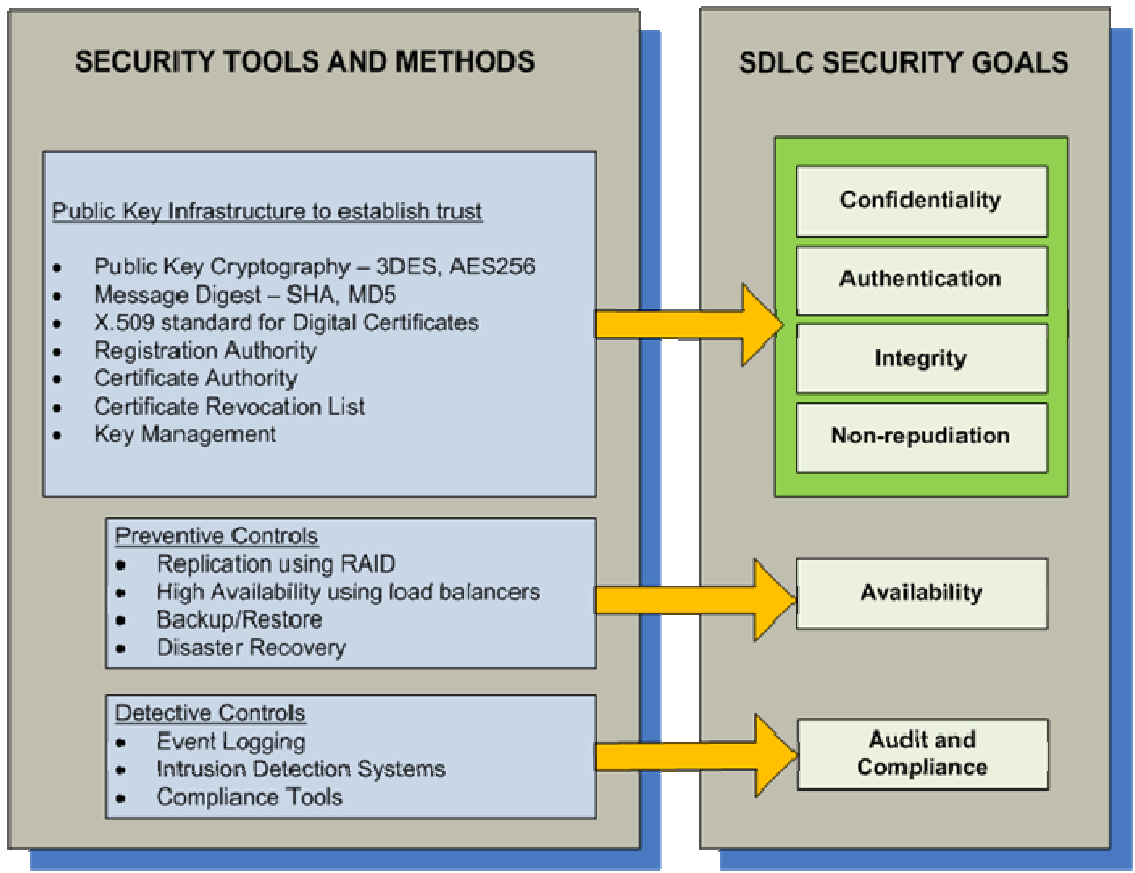

Figure 5.  Security in SDLC [23]

## 6. IDENTITY MANAGEMENT WORKFLOWS

The enhancement of the identity management workflows will address the problem of inconsistent user experience due to lack of end-to-end event driven user provisioning and de-provisioning of resources.  This section will describe the required workflows of the identity management system.  Workflows are a set of steps that define business processes.

As described earlier, PeopleSoft HCM/SA is the authoritative source of new identities and all updates to identities.  The authoritative source determines the new user's role to be one of the following: a student, an employee, faculty, or contractor.  The identity management system pulls the new user's information from PeopleSoft HCM/SA and creates an account for a user in the identity management system.  Specifically, IDM provisions resources to this user, based on predefined authorization rules related to the user's role.  For instance, if this user's role is an employee then IDM will provision Active Directory and Microsoft Exchange, and LDAP.  The IDM system is designed to automate user provisioning based on roles.  This process of workflow automation will be deployed to the non-production and production regions to ensure consistency of role-based provisioning and de-provisioning across all environments.

The three primary workflows that will be developed are:

- Provisioning workflow which will:
  1. Pull information from the authoritative source
  2. Create the user's account in the Identity Management system
  3. Based on the user's role, provision the user with the managed resources

- Identity update workflow which will:
    1. Pull information from the authoritative source
    2. Update the user's account in the identity management system
    3. Based on the type of modification needed, update the user's account on the managed resources side, and update access rights

- De-provisioning workflow which will:
    1. Pull information from the authoritative source
    2. De-provision the user from all of the managed resources

Figure 6, illustrates the flow of data from the authoritative source to the IDM system and finally to the managed resources. As depicted on the diagram, it is not only important to establish trust and to secure all the communication channels between all components, but also necessary to establish trust and security internally within each component.

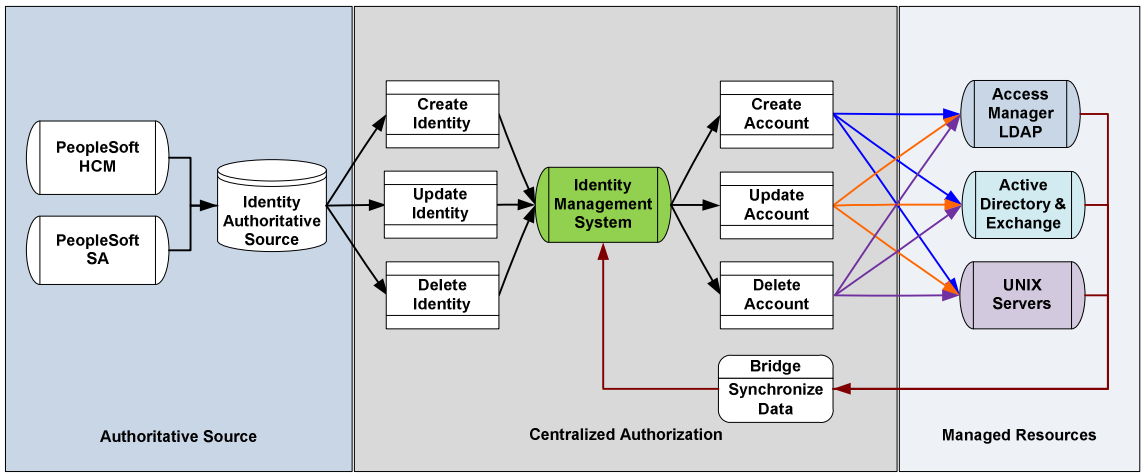

Figure 6. Identity provisioning and de-provisioning [24]

For instance, the IDM system component must configured as "root" to perform user administration tasks on the UNIX servers that it manages. It must be able to add accounts, delete accounts, change passwords for accounts, suspend accounts, restore accounts, and retrieve account data. Using SSH and its tunneling functionality is a countermeasure that should be used to manage this exposure. Internal to the IDM server and the UNIX server, is SSH client and SSH server, respectively. Older versions of SSH were vulnerable to SSH CRC-32 compensation attack detector so using the newest version of SSH is required. It is also important to enable the privilege separation feature, which is a mechanism to create a non-privileged environment for the sshd process to run in, should an intruder compromise SSH via a buffer overflow. (McClure, Scambray, & Kurtz, 2009)

Another requirement is that when the workflow adds users to Active Directory, the underlying security framework using ACLs within Active Directory must be in place. Developing workflows and enhancing the security of the identity management system requires the analysis of user input entry points, and process entry points. These entry points require security controls to preserve the integrity, confidentiality, and availability of the identity management infrastructure.

## 6.1. Cipher Attacks

Table 6 defines several common cipher attacks that must be taken into consideration into enhancing the security of the identity management infrastructure and the development of identity management workflows.

Table 6.  Risks in the identity management infrastructure. [25]

| Type | Initiation | Target System/s | Risks |
|------|-----------|-----------------|-------|
| User | Browser and on-line application process | Access Management System, PeopleSoft HCM/SA, Identity Management System | Injections<br>- SQL<br>- LDAP |
| User | Browser and on-line application process | PeopleSoft HCM/SA | Cross Site Scripting |
| User | Browser through the authentication Process | Access Management System, PeopleSoft HCM/SA, Identity Management System | Broken Authentication and Session Management |
| Process | Browser through the authentication Process | IDM, Access Management, Active Directory, Microsoft Exchange, UNIX Servers, Blackboard Learning System, PeopleSoft HCM/SA/Finance | Insecure Direct Object References |
| Process | Browser accessing a protected object | IDM, Access Management, Active Directory, Microsoft Exchange, UNIX Servers, Blackboard Learning System, PeopleSoft HCM/SA/Finance | Failure to restrict URL Access, and Insufficient Transport Layer Protection |
| Process | While a session is established | IDM, Access Management, Active Directory, Microsoft Exchange, UNIX Servers, Blackboard Learning System, PeopleSoft HCM/SA/Finance | Insecure Cryptographic Storage |

### 6.1.1. Injections

Further Injections attacks are used against programs that fail to check input data prior to processing them.  As a result, the data that is entered will influence the flow of program's execution [26].  Alpha's online application form is an internet facing application that is used to gather data about a prospective student.  The type of data entered by a prospective student includes PII data.  This is a point of entry into the internal applications within Alpha's environment.  From the perspective of the attacker, this form would be exploitable using SQL or LDAP injections, if the interpreter or program that is used to accept data is programmatically weak and can be manipulated into providing confidential data, or if it can be used to execute harmful commands.  This type of attack can compromise the integrity, availability, and confidentiality of ERP system - PeopleSoft HCM/SA.  For instance, the following LDAP query example, obtained from http://www.owasp.org/index.php/LDAP_injection site, can be compromised to access a user's password if the username is known: In this example the user is admin who may have privileges equivalent to cn=root.

```
String ldapSearchQuery = "(cn=" + $userName + ")";
System.out.println(ldapSearchQuery);
```

If the variable $userName is not validated, it could be possible accomplish LDAP injection, as follows:

1. If a user puts "*" on box search, the system may return all the usernames on the LDAP base
2. If a user puts "admin) (| (password = * ) )", it will generate the code bellow revealing admin's password ( cn = admin ) ( | (password = * ) )

To manage this risk, a parameterized API instead of an interpreter will be used to prevent injections. This will ensure that legitimate users of the system will gain credible information from the system and will be assured that the system they are entering their information into is not phishing for information.

### 6.1.2. Cross Site Scripting (XSS)

XSS flaws result from an application taking untrusted data. Input provided by one user into a program is output to another user. "It involves the inclusion of script code in the HTML content of a Web page, displayed by a user's browser."

Encountering cross site scripting will cause integrity, confidentiality, and availability in an environment. Once again, Alpha's online application form that is internet facing can be vulnerable to cross site scripting. There are a variety of ways to inject malicious scripts into Web pages. The need to use structured mechanisms that ensure separation of data and code will help prevent XSS. It is a better approach instead of relying on the developer to provide this mechanism in all places where user input can be displayed as Web output [28]. In addition, using an application firewall as a secondary layer of security will guard against XSS.

### 6.1.4. Broken Authentication and Session Management

At Alpha, the access management system provides forms-based authentication to the users. When authentication or session management is compromised, one negative result is identity theft. When this happens, not only the user is compromised, but also the credibility of Alpha Educational Institution on a business level which can affect revenues. Using the capabilities of the access management system to manage user passwords, and session tokens must be deployed as a preventive mechanism. Establishing trust on the server side instead of on the client side will ensure that attackers have not bypassed authentication by modifying values after the checks are done locally on the client machine. In addition, limiting the amount of failed authentication attempts will help prevent brute force attacks.

### 6.1.4. Insecure Direct Object References

Once a user has successfully authenticated, the next step is authorization. It is a good practice to have zero authorization as a default and provide only what is needed. Knowing the application well-enough to determine areas that should allow anonymous access, normal, privileged, and administrative is imperative to preventing insecure direct object references. Limiting the user's authorization to only what is needed can be controlled with the use of Access Control Lists (ACLs). Using indirect reference maps, avoiding the exposure of file names, external and internal URL's, and database keys is a sensible approach to mitigating this risk. Mitigating insecure direct references is not only directed for Alpha's clients (i.e. students, and faculty), but also to employees and contractors. When these preventive measures are implemented, it will preserve confidentiality, integrity, and availability of the data, as well as, the application infrastructure.

### 6.6.1. Failure to restrict URL Access, and Insufficient Transport Layer Protection

This vulnerability can be exploited when the application fails to check URL access rights every time it renders Web pages, protected links, and buttons. External to the application server is the network that is used to transport information from one component to another. An application that fails to implement preventive controls to secure the data being transported from one component to another is susceptible to cipher attacks.

For instance, when the prospective student is entering PII, the data being sent to the application server and into the database must be protected during transport. The data flows from the browser to the application portal which is located on the web server, from the web server to the application itself, and from the application to the database that is utilized by this application. This is shown on Figure 7. Implementing SSL between all the components, from the user's browser and to all the components will ensure that sensitive data can retain its confidentiality, and integrity [29].

Implementing SSL entails the use of signed digital certificates that are issued by a CA who verifies the authenticity of entities. In this case, the entities are the user's laptop, application portal which is on the web server, the application itself, and the database.

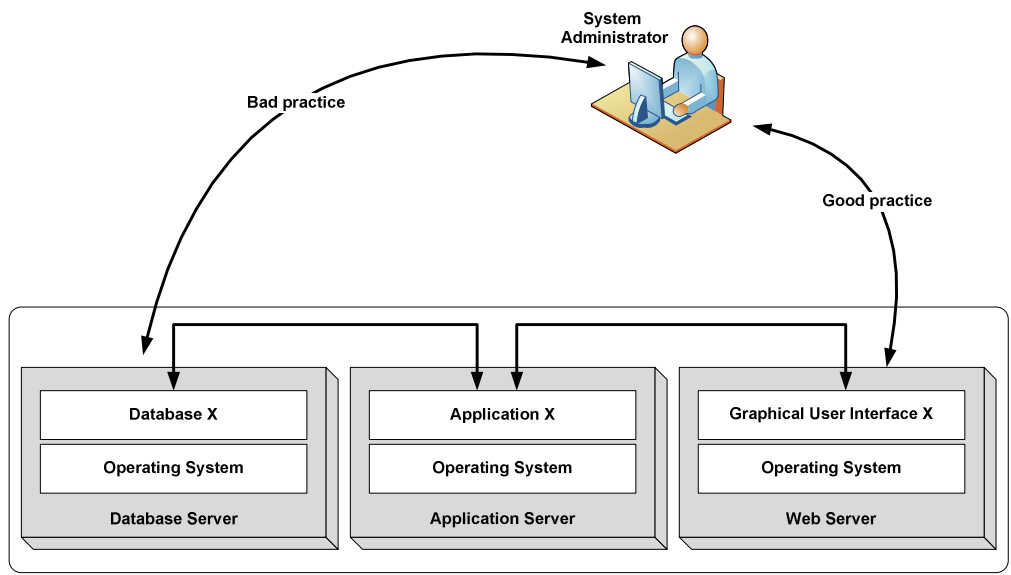

Figure 7. User interaction with the application [30]

### 6.6.1. Insecure Cryptographic Storage

The PCI standard states that credit card information should be encrypted in storage. Therefore, to comply with PCI standards, and with Alpha's security objectives, a method of encryption must be deployed to protect sensitive data including passwords. It's important to note that a typical information lifecycle entails data in storage, data in use, and data in transport as it can be moved, or backed up, and eventually deleted. Symmetric key algorithm (e.g. 3DES) is the recommended encryption method for the data in the Oracle databases that are utilized by PeopleSoft HCM/SA, including the identity and access management systems.

Understanding the different components of the identity management system helps ensure that an efficient encryption method is used so that functionality cannot be compromised. So a prudent approach would be that if the two components are communicating within the same server, then the use of symmetric or private key (i.e. 3DES) is recommended. If the two components communicating to each other are located in separate servers, then using public or asymmetric key (i.e. RSA or DSA) is recommended to ensure confidentiality, authenticity and non-repudiation.

For instance, authentication in a PKI environment shown on Figure 2 is recommended for all user sessions. The user is accessing the application portal that is secured by the access management system. RSA or DSA will be the encryption algorithm for all users' sessions. To recapitulate, all the entry points, namely, the online application form, the authentication page for accessing web applications, the administrative user interface used by Alpha's support staff must all be designed securely to prevent cipher attacks mentioned on Table 6. All the connection points between each component, shown on Figure 6, must be secured with SSL. When all the underlying security framework of the identity and access management is implemented, the three primary workflows that automate user provisioning and de-provisioning will function securely as expected.

# 7. CONCLUSION

Implementing the strategic initiatives described in this paper provides a foundation for centralizing security by utilizing the identity and access management systems. To summarize, defining consistent and repeatable processes enables online universities to understand the underlying practices and thought process involved when making access rights decisions. Refining the sequence of steps to achieve efficiency and automation is the fundamental requirement in workflow development. Establishing consistent and secure processes for end-to-end user provisioning and de-provisioning will result to secure, seamless, and consistent user experience. Using role-based access to provision resources to users is a way to ensure least privilege access by providing only the entitlements necessary to complete the tasks. Deploying single sign-on, fine-grained authorization, and delegated administration by using the access management system enables the organization to centralize access control. Enhancing the security of identity and access management systems means evaluating all the integrated components to ensure that they have all been configured securely using by incorporating PKI, database security, and secure application practices. Building security into these components means understanding the SDLC process, and the importance of software security which implies protecting the organization from cipher attacks. Overall, maintaining a secure organization means mitigating internal and external risks, providing security training and imparting the awareness of knowing how to consistently remain compliant to PCI and FERPA. Understanding the foundation of IAM will result to efficiency, and agility for managing growth through increased enrollments or future acquisitions. Realizing security synthesis in online universities means exemplifying the philosophy of creating an environment that fosters confidentiality, integrity, availability, and information assurance.

# REFERENCES

[1]     PCI Security Standards Council, (Oct. 2010) PCI DSS Quick Reference Guide Understanding the Payment Card Industry Data Security Standard Version 2.0, retrieved January 16, 2011 from https://www.pcisecuritystandards.org/documents/PCI%20SSC%20Quick%20Reference%20Guide.pdf.

[2]     US Department of Education, (2011), The Family Educational Rights and Privacy Act Guidance for Eligible Students, *FERPA General Guidance for Students,* retrieved March 5, 2011 from http://www2.ed.gov/policy/gen/guid/fpco/ferpa/for-eligible-students.pdf.

[3]     Modeled after "Figure 19-2 IBM Tivoli Identity Manager Relationships," (p. 587), Chapter 19: Identity Manager Scenarios, *Enterprise Security Architecture Using IBM Tivoli Security Solutions, IBM Rebooks Publication, SG24-6014-04*, 2007.

[4]     Table derived from "Figure 17-4 User/Role/Service/Group relationships," (p.532), Chapter 17: Identity Management, *Enterprise Security Architecture Using IBM Tivoli Security Solutions, IBM Rebooks Publication, SG24-6014-04*, 2007.

[5]     Table derived from "Table 17-3 User to Repository Mapping Role," (p. 535), Chapter 17: Identity Management, *Enterprise Security Architecture Using IBM Tivoli Security Solutions, IBM Rebooks Publication, SG24-6014-04*, 2007.

[6]     McCallister, E., Grance, T., & Scarfone, K. (Apr. 2010). Guide to Protecting the Confidentiality of Personally Identifiable Information (PII), retrieved Feb 6, 2011 from http://csrc.nist.gov/publications/nistpubs/800-122/sp800-122.pdf.

[7]     Raval, V., & Fichadia, A., (2007). *Risks, Controls, and Security: Concepts and Applications,* Hoboken, NJ: Wiley.

[8]     Harris, S. (2007*). CISSP All-in-One Exam Guide,* New York, NY: McGraw-Hill Osborne Media, p.729

[9]     Raval, V., & Fichadia, A., (2007). *Risks, Controls, and Security: Concepts and Applications,* Hoboken, NJ: Wiley.

[10]    Retrieved on February 6, 2011, Derived from "Figure 4-6 Basic DMZ Design," (p. 121), *Chapter 4: Common Security Architecture and Network Models* from http://www.redbooks.ibm.com/redbooks/pdfs/sg247581.pdf.

[11]    Aboba, B., Blunk, L., Vollbrecht, J., Carlson, J., & Levkowetz, H., (2004), Extensible Authentication Protocol (EAP), *Internet Engineering Task Force,* retrieved Feb 6, 2011 from http://www.ietf.org/rfc/rfc3748.txt.

[12]    Deacon,  A., & Hurst, R.,  (2007), RFC 5019: The Lightweight Online Certificate Status Protocol (OCSP) Profile for High-Volume Environments, retrieved February 6, 2011 from http://www.rfc-archive.org/getrfc.php?rfc=5019.

[13]    Originated from Table 5-1 Delegated administration roles in Access Manager, (p. 184), Chapter 5: Access Manager Core Components, *Enterprise Security Architecture Using IBM Tivoli Security Solutions, IBM Rebooks Publication, SG24-6014-04*, 2007.

[14]    Diagram derived from "Example SSL Configurations," retrieved February 6, 2011, from http://publib.boulder.ibm.com/infocenter/tivihelp/v2r1/index.jsp?topic=/com.ibm.itim.doc_5.0/cpt/cpt_ic_security_ssl_midware_setup.htm&resultof=%22SSL%22%20%22ssl%22%20%22Configuration%22%20%22configur%22.

[15]    Derived from "Figure 8.2 Two-tier versus three-tier application architecture," (p. 206), Raval, V., & Fichadia A. "Chapter 8: Application Primer" *Risks, Controls, and Security: Concepts and Applications,* Hoboken, NJ: Wiley, 2007, Print.

[16]    Derived from "Requirements for the Creation and the Maintenance of Student Data in Electronic Repositories," retrieved, March 12, 2011, from htttp://registrar.colorado.edu/sis_replacement/interfaces_and_reporting/docs/requirements_shared_data_respositories.pdf.

[17]    Derived from "Requirements for the Creation and the Maintenance of Student Data in Electronic Repositories," retrieved, March 12, 2011, from http://registrar.colorado.edu/sis_replacement/interfaces_and_reporting/docs/requirements_shared_data_respositories.pdf.

[18]    Raval, V., & Fichadia, A., (2007). *Risks, Controls, and Security: Concepts and Applications,* Hoboken, NJ: Wiley.

[19]    Raval, V., & Fichadia, A., (2007). *Risks, Controls, and Security: Concepts and Applications,* Hoboken, NJ: Wiley.

[20]    Raval, V., & Fichadia, A., (2007). *Risks, Controls, and Security: Concepts and Applications,* Hoboken, NJ: Wiley.

[21]    Raval, V., & Fichadia, A., (2007). *Risks, Controls, and Security: Concepts and Applications,* Hoboken, NJ: Wiley.

[22]    McBride, P., & Moser, E. P., (2000) Building Security into Your System--Not Bolting It On After the Damage Is Done, *Secure System Development Lifecycle (SDLC),* retrieved January 30, 2011 from http://www.lazarusalliance.com/horsewiki/images/3/38/Secure-System-Development-Life-Cycle.pdf.

[23]    Derived from "Figure 4 Security Goals, Methods, and Tools" McBride, P., & Moser, E. P., (2000) Building Security Into Your System--Not Bolting It On After the Damage Is Done. *Secure System Development Lifecycle (SDLC)*, retrieved January 30, 2011 from http://www.lazarusalliance.com/horsewiki/images/3/38/Secure-System-Development-Life-Cycle.pdf.

[24]    Derived from "Figure 19-2 IBM Tivoli Identity Manager Relationships," (p. 587), Chapter 19: Identity Manager Scenarios, *Enterprise Security Architecture Using IBM Tivoli Security Solutions, IBM Rebooks Publication, SG24-6014-04*, 2007.

[25]    This table was created based on information gathered from each description of OWASP Top 10 Application Security Risks – 2010, retrieved on March 6, 2011 from http://www.owasp.org/index.php/Top_10_2010-Main.

[26]    Stallings, W., Brown, L., Bauer, M. & Howard, M., (2008), *Computer Security: Principles and Practice*, Upper Saddle River, NJ: Pearson Prentice Hall.

[27]    Stallings, W., Brown, L., Bauer, M. & Howard, M., (2008), *Computer Security: Principles and Practice*, Upper Saddle River, NJ: Pearson Prentice Hall, p 398

[28]    Raval, V., & Fichadia, A., (2007). *Risks, Controls, and Security: Concepts and Applications,* Hoboken, NJ: Wiley.

[29]    Stallings, W., Brown, L., Bauer, M. & Howard, M., (2008), *Computer Security: Principles and Practice*, Upper Saddle River, NJ: Pearson Prentice Hall.

[30]    Derived from "Figure 8.2 Two-tier versus three-tier application architecture," (p. 206), Raval, V., & Fichadia A. "Chapter 8: Application Primer" *Risks, Controls, and Security: Concepts and Applications*. Hoboken, NJ: Wiley, 2007. Print.

**Authors**

**Maria Schuett** is a Certified Risk and Information Systems Control (CRISC) Security Consultant and a graduate student in Information Security and Assurance at Capella University. She has over 10 years of technical experience in information security and technology. She has delivered identity and access management solutions in the state government, health care, and insurance industries. She has written for her clients whitepapers about infrastructure resilience, and service management surrounding access control. Her manuscript about Reduced Sign-On is included in the Encyclopedia of Information Assurance (http://isbn.nu/9781420066203/).

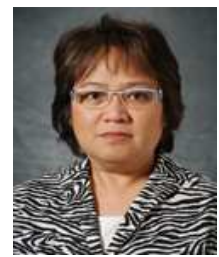

**Syed (Shawon) M. Rahman** is an assistant professor in the Department of Computer Science and Engineering at the University of Hawaii-Hilo and an adjunct faculty of information Technology, information assurance and security at the Capella University. Dr. Rahman's research interests include software engineering education, data visualization, information assurance and security, web accessibility, and software testing and quality assurance. He has published more than 65 peer-reviewed papers. He is a member of many professional organizations including ACM, ASEE, ASQ, IEEE, and UPE.

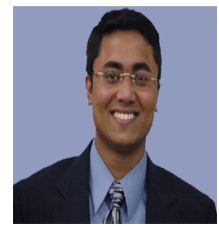